\documentclass[12pt]{spieman}  % 12pt font required by SPIE;
\usepackage{aas_macros}
\usepackage{amsmath,amsfonts,amssymb}
\usepackage{graphicx}
\usepackage{setspace}
\usepackage[subfigure]{tocloft}
\usepackage{subfigure}
\usepackage{comment}

\usepackage{bm}
\usepackage[dvipsnames]{xcolor}
\usepackage{verbatim}
\usepackage{graphicx}

\title{The Hydrogen Intensity and Real-time Analysis eXperiment: 256-Element Array Status and Overview}

\author[a,b,c,*]{Devin~Crichton}
\author[b,c]{Moumita~Aich}
\author[d,a]{Adam~Amara}
\author[e,f]{Kevin~Bandura}
\author[g,h,i]{Bruce~A.~Bassett}
\author[j,k]{Carlos~Bengaly}
\author[a]{Pascale~Berner}
\author[l]{Shruti~Bhatporia}
\author[m,n,b,o]{Martin~Bucher}
\author[p]{Tzu-Ching~Chang}
% ...
\affil[a]{Institute for Particle Physics and Astrophysics, ETH Z\"urich, Z\"urich, Switzerland}
\affil[b]{School of Mathematics, Statistics, \& Computer Science, University of KwaZulu-Natal, Durban, South Africa}
\affil[c]{Astrophysics Research Centre, University of KwaZulu-Natal, Westville Campus, Durban 4000, South Africa}
\affil[d]{Institute of Cosmology \& Gravitation, University of Portsmouth, Dennis Sciama Building, Burnaby Road, Portsmouth PO1 3FX, UK}
\affil[e]{LCSEE, West Virginia University, Morgantown, WV 26506, USA}
\affil[f]{Center for Gravitational Waves and Cosmology, West Virginia University, Morgantown, WV 26505, USA}
\affil[g]{African Institute for Mathematical Sciences, 6 Melrose Road, Muizenberg, 7945, South Africa}
\affil[h]{Department of Maths and Applied Maths, University of Cape Town, Rondebosch, Cape Town, 7700, South Africa}
\affil[i]{South African Astronomical Observatory, Observatory, Cape Town, 7925, South Africa}
\affil[j]{D\'epartement de Physique Th\'eorique, Universit\'e de Gen\`eve, 1211 Gen\'eve 4, Switzerland}
\affil[k]{Observat\'orio Nacional, 20921-400, Rio de Janeiro - RJ, Brazil}
\affil[l]{High Energy Physics, Cosmology \& Astrophysics Theory Group, University of Cape Town, Cape Town, South Africa}
\affil[m]{Universit\'e de Paris, CNRS, Astroparticule et Cosmologie, F-75013 Paris, France}
\affil[n]{LPENS, Ecole Normale Sup\'erieure, 75005 Paris, France}
\affil[o]{NITheCS, Stellenbosch, South Africa}
\affil[p]{NASA Jet Propulsion Laboratory, Californita Institute of Technology, Pasadena, CA, USA}
% ...

\cftpagenumbersoff{figure}
\cftpagenumbersoff{table} 

\usepackage{xspace}
\def\chime{CHIME}
\def\hartrao{HartRAO}
\def\hirax{HIRAX}
\def\rfof{RFoF}
\def\hone{H{\sc i}}

\begin{document} 
\maketitle

\begin{abstract}
The Hydrogen Intensity and Real-time Analysis eXperiment (\hirax) is a radio interferometer 
array currently in development, with an initial 256-element array to be deployed at the South African Radio Astronomy Observatory (SARAO) Square Kilometer Array (SKA) 
site in South Africa. Each of the 6~m, $f/0.23$ dishes will be instrumented with dual-polarisation feeds operating 
over a frequency range of 400--800~MHz. Through intensity mapping of the 21~cm emission line of neutral 
hydrogen, \hirax\ will provide a cosmological survey of the distribution of large-scale structure
over the redshift range of $0.775 < z < 2.55$ over $\sim$15,000 square degrees of the southern sky. The statistical power of such a survey 
is sufficient to produce $\sim$7 percent constraints on the dark energy equation of state
parameter when combined with measurements from the \textit{Planck} satellite. Additionally,
\hirax\ will provide a highly competitive platform for radio transient and HI absorber science 
while enabling a multitude of cross-correlation studies. In this paper, we describe 
the science goals of the experiment, overview of the design and status of the sub-components of 
the telescope system, and describe the expected performance of the initial 256-element array as well 
as the planned future expansion to the final, 1024-element array.
\end{abstract}

% Include a list of up to six keywords after the abstract

% Include email contact information for corresponding author
{\noindent \footnotesize\textbf{*}Devin Crichton,  \linkable{dcrichton@phys.ethz.ch} }

\begin{spacing}{1}
%\begin{spacing}{2}   % use double spacing for rest of manuscript

\section{Introduction}\label{s:intro}

In recent years, the statistical distribution of large-scale structure in the late-time universe 
has been measured with increased precision. Current state-of-the-art measurements primarily
come from baryon acoustic oscillation (BAO) studies from galaxy and Lyman-$\alpha$ surveys as well 
as weak lensing probes\cite{2017MNRAS.470.2617A,2021PhRvD.103h3533A,2013JCAP...04..026S,2021arXiv210513549D,2014MNRAS.441.2725F,2021A&A...646A.140H,2019PASJ...71...43H}.
However, tomographic measurements in the visible/infrared bands become increasingly difficult beyond
$z\sim1$, requiring complex optical systems and long integration times.
Intensity mapping with the 21~cm (1420.4~MHz) line of neutral hydrogen has the potential to push 
beyond these limitations, with observations in the range of $0 < z \lesssim 30$ being
 achievable with current and planned instruments. The abundance of hydrogen and the optically thin nature of 
the emission will enable detailed tomographic measurements of the distribution of matter over large, 
cosmological volumes\cite{2008PhRvL.100i1303C,2015ApJ...803...21B,2017arXiv170909066K}. 
In the post-reionisation epoch, by observing the BAO 
feature in the matter power spectrum, intensity mapping enables cosmological measurements that can break parameter 
degeneracies in current combinations of early-universe and late-time measurements, as well 
as constrain the dynamical nature of dark energy over the transition from matter-dominated to 
dark-energy dominated expansion.

Modern advances in computer and communication hardware as well as the relatively simple, low-frequency receivers required for these 
observations have motivated the development of large, $\sim$100--1000 element interferometer arrays such as 
\chime\cite{2014SPIE.9145E..22B}, CHORD\cite{2019clrp.2020...28V}, HERA\cite{2017PASP..129d5001D}, PUMA\cite{2019BAAS...51g..53S}, and Tianlai\cite{2012IJMPS..12..256C}.  Such telescopes focus their sensitivity on cosmologically relevant 
angular scales, favouring compact arrays with ``redundant'' configurations, where the constituent radio
telescopes are placed with regular spacing.  This type of array architecture additionally
provides a natural platform for the detection and monitoring of radio transients, 
with wide fields of view, fast mapping speeds, and the real-time processing power required for fast 
radio burst (FRB) detection and pulsar searches, as demonstrated by CHIME in, e.g. Refs.~\citenum{2019Natur.566..230C,2019Natur.566..235C,2019ApJ...885L..24C,2021arXiv210604352T,2020MNRAS.496.2836N}.

\begin{figure}
    \centering
    \includegraphics[width=\textwidth]{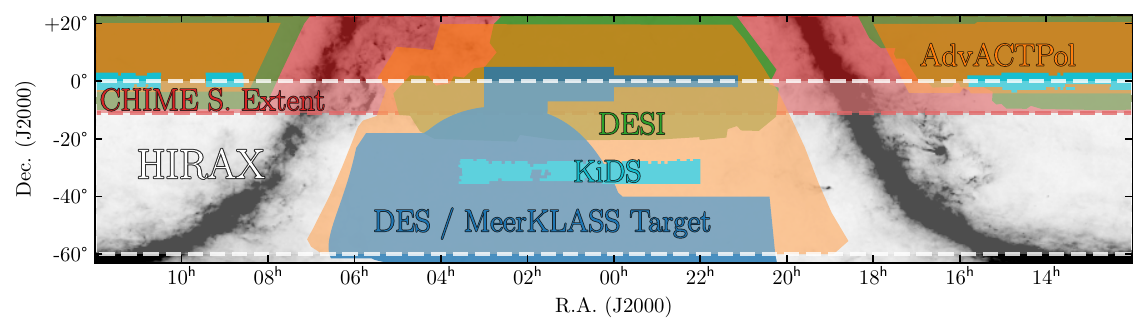}
    \caption{\hirax\ will observe $\sim$15,000~deg$^2$ of the southern
      sky, shown by the region enclosed by the white dashed lines. The survey footprint overlaps many existing and forthcoming surveys such as those of 
      \chime\cite{2014SPIE.9145E..22B}, DES\cite{2021arXiv210513549D}, KiDS\cite{2021A&A...646A.140H}, DESI\cite{2016arXiv161100036D} and Advanced ACTPol\cite{2016JLTP..184..772H}.
      The entirety of the sky area shown is contained within the Rubin/LSST\cite{2009arXiv0912.0201L} survey footprint and is accessible to the SKA\cite{2015aska.confE.174B} and its precursors. The background colorscale shows Galactic emission from the \textit{Planck} 353~GHz map.}
    \label{fig:survey_overlap}
\end{figure}

The Hydrogen Intensity and Real-time Analysis eXperiment (\hirax)\cite{2016SPIE.9906E..5XN} is a
new radio interferometer array that will employ a compact, redundant layout. \hirax\ is currently funded for the construction of an initial 256-element array (\hirax-256) 
that will comprise a subset of the planned 1024-element array.
The telescope will conduct a large 21~cm intensity mapping survey, targeting cosmological constraints on the dark energy equation-of-state parameters, while additionally operating as a platform for transient and HI absorber searches.
The cosmological survey requires careful control of systematics due to significant risk of contamination by residuals from foregrounds, which are up to six orders of magnitude brighter in power.
The radio transient detection pipeline for \hirax\ will focus on searches for FRBs and millisecond pulsars. FRBs are currently under intense study since the nature of their emission processes is not yet understood. \hirax-256 is expected to detect multiple FRBs per week\footnote{Based on the all-sky FRB rate estimate from Ref.~\citenum{2021arXiv210604352T}, scaled to the \hirax\ field of view, noting that \hirax-256 has a similar collecting area to CHIME.}
and future proposed \hirax\ outrigger sites with very long baselines ($\sim$1000~km) will aid in the localisation of FRB detections to within 0.1 arcseconds. Pulsar searches will be performed on beam-formed data at full baseband, allowing for searches over a wide parameter space. Additionally, the flexible digital backend will allow for the monitoring of known pulsars (aiding existing pulsar timing studies) as well as resampling the full baseband data to $\sim$3~kHz spectral resolution, 
which will be used to conduct a blind \hone\ absorber search.
Such a search will provide an informative accounting of the state of cold gas in galaxies out to $z\sim 2.5$, encompassing the peak of the global star-formation rate density at $z\sim 2$. Finally, as shown in Figure~\ref{fig:survey_overlap}, the $-60^\circ \lesssim \delta \lesssim 0^\circ$ declination range of the \hirax\ survey has significant overlap with a range of 
other large surveys of the southern and equatorial sky, thus enabling a wide range of potential cross-correlation studies. These include optical/infrared galaxy and lensing surveys from eBOSS\cite{2017AJ....154...28B}, DES\cite{2021arXiv210513549D}, KiDS\cite{2021A&A...646A.140H}, HSC\cite{hscsurvey}, DESI\cite{2016arXiv161100036D}, Rubin/LSST\cite{2009arXiv0912.0201L}, Euclid\cite{2018LRR....21....2A} and Roman\cite{2019arXiv190205569A} as well as ground based cosmic microwave background (CMB) surveys from Advanced ACTPol\cite{2016JLTP..184..772H}, SPT-3G\cite{2014SPIE.9153E..1PB} and the Simons Observatory\cite{2019JCAP...02..056A}. 

These science goals are highly synergistic with those of the SKA\cite{2015aska.confE.174B} and its precursors. For example, the combined MeerKLASS\cite{2017arXiv170906099S} and
\hirax\ intensity mapping surveys will have the potential to produce a cosmological analysis over a redshift range
of $0 \lesssim z \lesssim 2.5$, with contrasting systematic challenges from the differing observing modes and instruments. 
HIRAX will share significant overlap with the proposed SKA1-MID intensity mapping survey\cite{2015aska.confE..19S,2020PASA...37....7S} in terms of sky area ($\Omega_{\rm SKA1-MID} \approx$ 20,000 deg$^2$; in Figure~\ref{fig:survey_overlap} this covers most of the southern sky) and redshift range ($0<z_{\rm SKA1-MID}<3.06$), which will be useful for validation and testing systematics. At the same time, HIRAX will be complementary to the SKA1-MID single dish and interferometer intensity mapping surveys in terms of angular scales probed, which leads to complementary scientific constraints as discussed in Section~\ref{s:analysis}.
The blind \hone\ absorber surveys with \hirax\ will extend observations beyond the redshift limits of ongoing surveys, such as the MeerKAT Absorption Line Survey (MALS\cite{mals2016}), to investigate the relationship between cold atomic gas and the evolution of star formation and AGN activity. 

In this work, we present the current status and design of \hirax-256, as well as the analysis work that has been done to inform these design choices. In Section~\ref{s:instrument} we introduce the instrument and briefly outline the various sub-components. Section~\ref{s:analysis}
overviews the forecasted cosmological constraints and current analysis methodologies. In Section~\ref{s:requirements} we describe 
how forecasts and analysis of simulated data have constrained the design of the instrument, and finally in 
Section~\ref{s:conclusions} we discuss the current status of the project.

\section{Instrument and Subsystems}\label{s:instrument}

\subsection{Overview}

\begin{figure}
    \centering
    \includegraphics[width=0.9\linewidth]{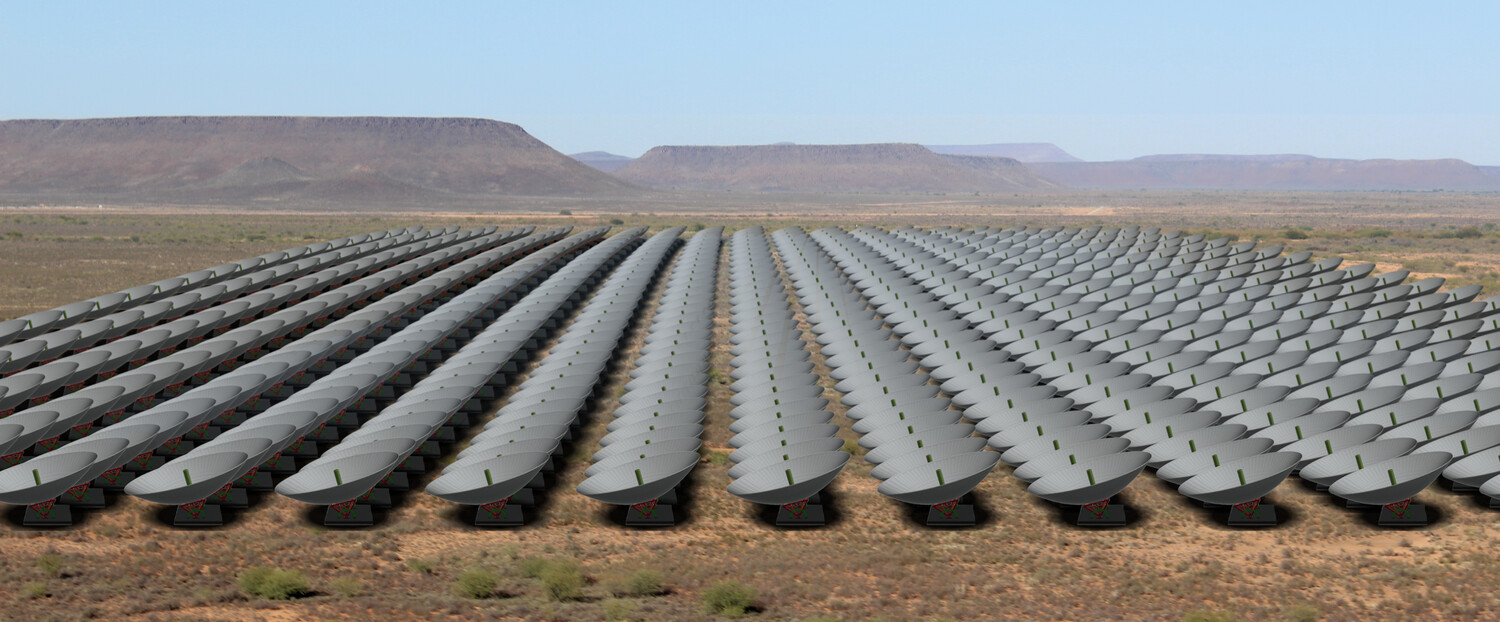}
    \caption{Artist's impression of the \hirax\ array in the Karoo desert.}
    \label{fig:hirax_karoo}
\end{figure}

The overall design of \hirax\ is driven by optimizing sensitivity to
the restricted range of angular scales corresponding to the BAO
signature.  The BAO science
goals thus naturally dictate the high-level instrument configuration.  The
redshift range $0.775 < z < 2.55$, which captures the onset of the influence of
dark energy on the expansion rate at $z \sim 2$, sets the observing frequency range of
400--800~MHz.  The 150~Mpc BAO characteristic length\cite{2007ApJ...664..660E} corresponds to
angular scales of $3^\circ$--$1.3^\circ$ at 400--800~MHz, thus setting
a minimum baseline length range of 15--60~m.  Along the line of sight,
the BAO characteristic length translates into frequency scales of
12--20~MHz, thus setting an absolute minimum requirement of $\sim$100
frequency channels over 400--800~MHz.  Finally, because the BAO signal
is faint ($\sim$0.1~mK), the instrument must have a large collecting
area and low system temperature.  The basic \hirax\ instrument
parameters are summarized in Table~\ref{t:inst_summary}, and a
rendering of the array is shown in Figure~\ref{fig:hirax_karoo}.
The \hirax\ array, which will initially consist of 256 elements, will
be installed as a guest instrument at the South African Radio Astronomy Observatory (SARAO)
site in the Karoo Desert, South Africa, at $30^\circ 41' 47''$~S,
$21^\circ 34' 20''$~E, approximately 12~km from the core of MeerKAT.
As a guest instrument, HIRAX will benefit from shared power and telecommunications
infrastructure provided under agreement with SARAO. From this site, \hirax\ will be able to access
$\sim$15,000~deg$^2$ of the southern sky, as illustrated in
Figure~\ref{fig:survey_overlap}.

To keep the total data volume at a manageable level, we will take
advantage of the redundant configuration of \hirax\ to average
visibilities within groups of nominally identical baselines.
A continuous few-day buffer of the full set of raw visibilities will be stored while the 
calibration and averaging process is performed, after which the raw
data will be deleted.  This averaging process is highly sensitive to
differences in telescope pairs, since even small mismatches may
couple with Galactic emission, which is up to $10^3$ times greater in amplitude than
the BAO signal, subsequently offsetting the calibration and
introducing systematic errors in the observations.  Redundant radio
telescope arrays that average visibilities early in this way, such as \hirax, therefore operate in a unique regime
that demands precise {\it control}---not merely
characterization---over possible departures from redundancy. The
methodology for deriving the \hirax\ instrument specifications is
discussed in further detail in \S\ref{s:requirements}.

 The degree to which baselines are redundant and the angular scales over which that redundancy is distributed are important considerations in determining the optimal \hirax\ array layout. Evaluating array layouts that strike a balance between redundancy and more uniform $uv$-plane coverage\cite{2016ApJ...826..181D} is an active area of study. 
 Key factors favouring maximal redundancy include reduced raw data rate and ease of redundant calibration, whereas key factors favouring less than maximal redundancy include the array chromatic response, taking into account dish-feed and dish-dish coupling across the array, and the impact of instrument errors that cause systematic departures from redundancy.

\begin{figure}
    \centering
    \includegraphics[width=0.9\linewidth]{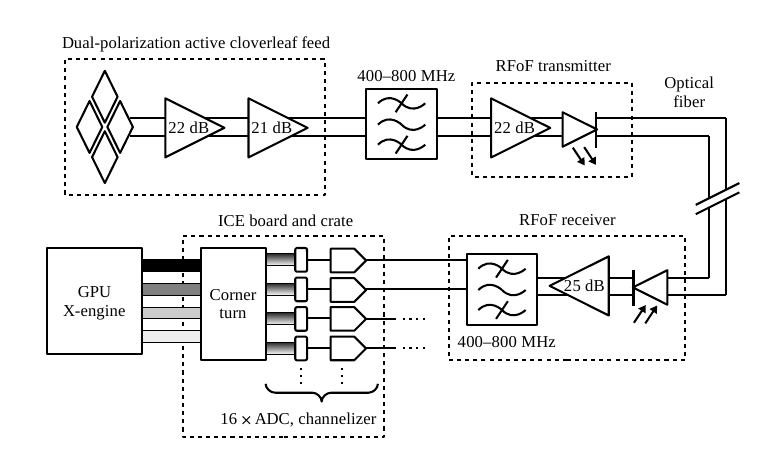}
    \caption{\hirax\ signal chain schematic.  In each dish, an active
      dual-polarization feed receives and amplifies incoming radio
      signals.  The signals are filtered, further amplified, and
      converted to optical light with an \rfof\ transmitter at the
      dish focus.  Optical fibre transports the signals to an
      \rfof\ receiver, which converts the signals back into RF and
      applies additional amplification and filtering before
      digitization.  An ICE-based system performs the digitization,
      channelizing, and corner turn.  The data are then passed to the
      GPU X-engine for correlation.}
    \label{fig:sig_schem}
\end{figure}

Figure~\ref{fig:sig_schem} illustrates the \hirax\ signal chain.  At
the focus of each telescope dish, the incoming radio signals are
received and amplified by a dual-polarization active feed, band
limited to 400--800~MHz, and converted to optical light with an RF
over optical fibre (\rfof) transmitter.  The optical signals from all
dishes are transmitted to a central processing hub and converted back
into RF using \rfof\ receivers, passed through a second stage of
400--800~MHz filtering, and then digitized.  \hirax\ employs ICE
boards~\cite{2016JAI.....541005B} for digitizing and channelizing the
RF inputs, and performing the corner turn operation before the data
are passed to the GPU X-engine for correlation.

\begin{table}
\centering
\begin{tabular}[pos]{ll}
\hline
{\bf Parameter} & {\bf Value} \\
\hline
Number of dishes & 256 \\
Dish diameter & 6~m \\
Dish focal ratio & 0.23 \\
Collecting area & 7200~m$^2$ \\
Frequency range & 400--800~MHz \\
Frequency resolution & 1024 channels, 390~kHz \\
Field of view & $5^\circ$--$10^\circ$ \\
Resolution & $0.2^\circ$--$0.4^\circ$ \\
Target system temperature & 50~K \\
\hline
\end{tabular}
\vskip 5pt
\caption{\hirax\ instrument parameters for the initial 256-element array.}\label{t:inst_summary}
\end{table}

\subsection{Dishes and Mounting Structure}\label{s:dish}

Each \hirax\ telescope will have a parabolic reflector with a diameter
of 6~m and focal ratio of $f/0.23$.  The dish size was chosen to
maximize collecting area over the minimum baseline lengths, and the
low focal ratio allows the entire receiver support structure to sit
below the aperture plane, thus reducing cross-talk between neighboring
dishes.  Each dish will be supported by a mounting structure that is
stationary in azimuth and that can be manually repointed in zenith
angle over a range of $\pm30^\circ$ to incrementally build up sky
coverage.  The \hirax\ redundancy requirements have strict
implications for the allowed tolerances on the dish surface and the
mount alignment, which impact the on-sky beam shape and pointing,
respectively.  In general, the requirements on dish and mount {\it
  precision} are far more stringent than the requirements on {\it
  accuracy}.  The precision requirements are discussed in further detail in~\S\ref{s:requirements}.

Dish fabrication methods using composite materials with an embedded
reflective layer are well suited for meeting the \hirax\ requirements
while being cost effective.  Fabricating all of the dishes using a
small number of molds ensures that variations across the array are
kept to a minimum.  Composite materials also have high
strength-to-weight ratios and are therefore excellent candidates for
minimizing gravitational distortion of the dish surfaces as the
telescopes are repointed.  One prototype composite dish has been
installed at the \hirax\ test site at the Hartebeesthoek Radio
Astronomy Observatory (\hartrao), and other prototyping efforts are
currently underway.  Further details about the \hirax\ prototype
dishes and design are available in Ref.~\citenum{2021SPIE11445E..5OS}.

\subsection{Dual-Polarization Cloverleaf Feed}

\hirax\ will use a dual-polarization cloverleaf feed based on the
design that was developed for \chime\cite{2017arXiv170808521D, 2021SPIE11445E..5OS}.  In
contrast to the \chime\ feed, which is passive, the \hirax\ feed is
active and provides on-board amplification with a powered balun. Each polarization of the \hirax\ feed consists of a balun that uses an Avago MGA-16116 dual amplifier, providing a gain of 22 dB, and the difference between the outputs are amplified by 21 dB using a Mini-Circuits PSA4-5043+. Each feed is mounted inside a cylindrical can, which circularizes the beam and helps reduce cross-talk.  Each feed and can assembly will be supported at the dish focus with a radio-transparent central column that extends upward from the vertex.  The column will provide environmental protection by fully enclosing the feed assembly, and the cables will be routed along the boresight axis to minimize their impact on beam asymmetry and sidelobe levels. Measurements of the noise performance of the HIRAX feed and its repeatability is described in Ref.~\citenum{2021arXiv210106337K}.

\subsection{Radio Frequency over Fibre System}

\hirax\ will make use of a purpose-built \rfof\ system to transmit the
received radio frequency signals to the digital backend for processing.
Signals received by the \hirax\ feed are band limited to 400--800~MHz
and passed through an additional 22~dB amplification stage before
being intensity modulated on an optical carrier at the focus using an
\rfof\ transmitter.  Optical fibres carry these signals to a central
processing hub, where \rfof\ receivers convert the signals back into
RF.  The RF signals are subsequently amplified and filtered again
(25~dB gain, 400--800~MHz) before being passed to the digitizer.  For
the long ($\sim$1~km) length of the \hirax\ cable runs from the telescopes to the
central hub, especially looking ahead to the planned 1024 element array,
the combination of \rfof\ electronics and optical fibres provide a
more cost-effective solution than copper coaxial cables.  The
\rfof\ transmitter and receiver design is based on technology that was
developed by \chime\cite{2013JInst...810003M} (although ultimately
unused in the final experimental configuration).  The transmitter
contains a laser diode (AGX Technologies, FPMR3 series\footnote{\url{http://www.agxtech.com/PDF/FPMR3+w+Isolator+Rev1.9.pdf}}) that is intensity modulated by the incoming RF signal, and the receiver
contains a photodetector (AGX Technologies, PPDD series\footnote{\url{http://www.agxtech.com/PDF/PPDA+R5.6.pdf}}) that converts
the transmitted optical signal into RF.

\subsection{ICE-based F-Engine}

The digital backend for \hirax\ consists of an FX correlator, with the F-engine comprised of
an ICE-based digitisation and channelisation system. The F-engine for \hirax-256 will be made up of 32 ICE boards, each processing 16 inputs from 8 dual-polarisation feeds. These ICE boards are contained in 
two ICE crates, each hosting 16 boards. The system for \hirax\  will match that of \chime\ which is described in more detail in Refs.~\citenum{2016JAI.....541004B,2016JAI.....541005B}. Briefly, the ICE boards use custom FPGA-based electronics to digitize the incoming signals from the RFoF system
at a precision of 8~bits\footnote{7.2 effective number of bits (ENOB), K. Bandura, Private Communication.} over the full 400~MHz of bandwidth. For the channelisation step, these digitised
signals are processed by a polyphase filter bank and FFT based pipeline, producing the 1024 frequency channels 
(390~kHz wide) that are passed to the correlator. This system also performs the real-time corner-turn operation,
which arranges the outgoing data such that each node of the X-engine receives data streams from all inputs
over the subset of the bandwidth to be processed by that node. 
This corner-turn is performed in multiple stages with data reshuffling occurring via communication within ICE boards, between ICE boards mounted in the same ICE crate, and finally between ICE crates. After this stage, the outgoing data streams are transferred to the GPU-based X-Engine over a dedicated 40~Gbps network. For the 1024 element array, this system will be scaled up to 128 ICE boards and 8 ICE crates. For \hirax-256, this system will process and forward data to the correlator at a rate of 1.65~Tbps, with the 1024 element system processing 6.6~Tbps.

\subsection{GPU Correlator}

The \hirax-256 X-engine is a dense, GPU-based system consisting of 8~nodes, each processing 128 channels or
50~MHz of bandwidth using a pair of NVIDIA A40 GPUs. These nodes will perform full, $N^2,$ correlation of the incoming data streams from 512 inputs (two orthogonal polarizations per dish). The full specifications of each of these correlator nodes
is shown in Table~\ref{t:correlator_node}.

\begin{table}[ht]
\centering
\begin{tabular}[pos]{ll}
    \hline
    {\bf Parameters} & {\bf Value} \\
    \hline
    Motherboard & GIGABYTE G482-Z52  \\
    Processor & 2 $\times$ AMD EPYC™ 7452, 32 cores each  \\
    RAM  & 1TB \\
    GPU & 2 $\times$ NVIDIA A40, PCIe~4.0 \\
    F-Engine Network & 4 $\times$ SILICOM PE31640G2QI71-QX4  - 2 $\times$ 40~Gbps \\
    Outgoing Data Network  & 2$\times$25~Gbps  \\
    \hline
\end{tabular}
\vskip 5pt
\caption{\hirax-256\ correlator node specifications. The \hirax-256\ X-engine is comprised of eight these nodes.}\label{t:correlator_node}
\end{table}

This system will produce raw visibility data for each baseline with an integration time of $\sim$10s. Additional outputs from
these nodes will include formed beams for the FRB and \hone\ absorber search pipelines as well as for the pulsar search
and monitoring systems, described in more detail in Section~\ref{sec:beamforming}.  This system represents a very dense and powerful correlator, processing 1.6~Tbps of data in a single rack, eight times more bandwidth per node than the CHIME X-engine~\cite{2015arXiv150306202D}. This has largely been enabled by making use of the increased I/O performance of PCIe 4.0 for both the GPU and network
card interconnects. The X-engine makes use of the \texttt{kotekan}\cite{ 2015arXiv150306189R,2015arXiv150306203K} software also used by \chime\ and CHORD. The correlator will be scaled up for the 1024-element array, potentially with an even denser layout depending on hardware developments.

\subsection{On-site Science Data Processing}

Both the visibility data and the beam-formed data from the X-engine will be sent to the on-site science data processing (SDP) system. This system will handle the initial calibration and baseline stacking operations for the cosmological analysis, produce data products for the \hone\ absorber studies and perform the real-time transient searches. Additionally, it will manage and accumulate incoming housekeeping data streams for later use. Reduced data products will be synchronised to off-site computing centres for further processing and long-term storage.

\section{Data Analysis Methodology}\label{s:analysis}

\subsection{Cosmological Survey}\label{s:survey}

\hirax\ will operate as a survey instrument to map the large volumes required for precise measurements of the BAO signal. The 21~cm cosmology science goals largely drive the system requirements (\S\ref{s:requirements}) for the array as a whole and the constituent telescopes. Here, we begin by outlining the top-level survey parameters that are required to meet the \hirax\ BAO science goals.

\begin{figure}[ht]
    \includegraphics[width=\textwidth]{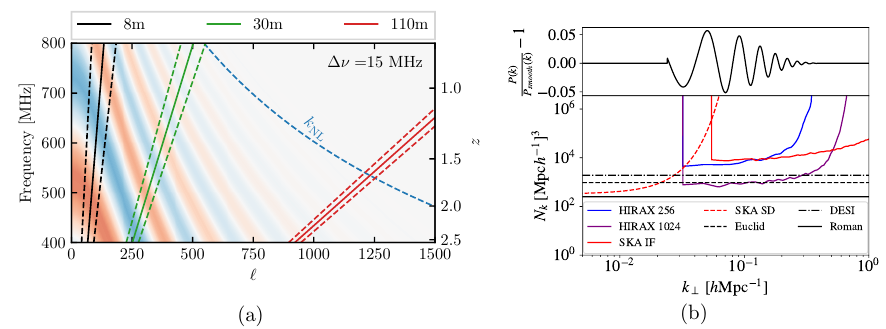}
    \caption{
    (a) Angular scales accessible to \hirax\ baselines as a function of frequency. The background colour scale represents the BAO feature in the 21~cm power spectrum assuming line-of-sight modes, $k_{\parallel},$ corresponding to correlations over 15~MHz in frequency. The blue curve shows scales smaller than which non-linear effects become important based on the approximate $k_{\rm NL}$ cut-off of $0.14$~Mpc$^{-1}$ from Ref.~\citenum{2015ApJ...803...21B} with the redshift scaling described in Ref.~\citenum{2003MNRAS.341.1311S}. These modes are excluded from the fiducial cosmological analysis as they are more difficult to map to the underlying cosmological parameters. This figure demonstrates the importance of \hirax's short baselines, and hence compact layout, for measuring the BAO feature in the linear power spectrum.
    (b) Comparison of \hirax's power spectrum sensitivity as a function of $k_{\perp}$ with that of SKA-MID surveys, in single dish (SD) and interferometer (IF) mode, shown alongside shot noise estimates from upcoming spectroscopic galaxy surveys from DESI, Euclid and Roman. The upper panel shows the BAO feature in the power spectrum as a function of $k_{\perp}$ at $z=1.2$. Differences in sensitivity due to different survey areas, and hence cosmic variance contributions, are not accounted for here.
    \label{fig:HIRAX_baselines_noise} 
    }

\end{figure}

To reduce cosmic variance, \hirax\ will survey $\sim$15,000 deg$^2$ from declinations
$-60^\circ \lesssim \delta \lesssim 0^\circ$ (Figure~\ref{fig:survey_overlap}) over a four year period with an observing efficiency of $\sim$50\%. The amount of time per fixed-elevation pointing of the array will be chosen to deliver a survey sensitivity uniform in declination across the mapped sky area to within 1.5\%. To probe the 21~cm power spectrum on BAO scales requires sensitivity to radial wavenumbers in the range $0.03 < k_\parallel < 0.2$~Mpc$^{-1},$ which can be achieved with the specified bandwidth and channelisation, and sensitivity to transverse wavenumbers in the range $0.05 < k_\perp < 0.16$~Mpc$^{-1},$ which can be achieved with baseline separations covering a minimum range of 8~m to 110~m, as can be seen in Figure~\ref{fig:HIRAX_baselines_noise}(a). The \hirax\ survey sensitivity is shown in Figure~\ref{fig:HIRAX_baselines_noise}(b), which demonstrates that \hirax\ is optimised to measure BAO scales where it is competitive with other planned intensity mapping and galaxy redshift surveys. For more details on the calculation of these sensitivity estimates, see Refs.~\citenum{2015ApJ...803...21B,2017PhRvD..96d3515A}.

The angular scales probed by HIRAX, targeted at constraining dark energy by making use of the BAO feature, complement the angular scales and thus the science probed by the SKA1-MID intensity mapping surveys as can be seen in Figure~\ref{fig:HIRAX_baselines_noise}(b). The SKA1-MID single dish survey will probe very large angular scales and be sensitive to ultra-large scale features such as primordial non-Gaussianity and general relativistic corrections to cosmological observables,\cite{Camera:2015NF} whereas the SKA1-MID interferometer survey will be sensitive to smaller angular scales and thereby constrain the density and bias of neutral hydrogen on small scales.

\subsubsection{Fisher Forecasts}\label{s:fisher}

\begin{figure}
    \centering
    \includegraphics[width=\textwidth]{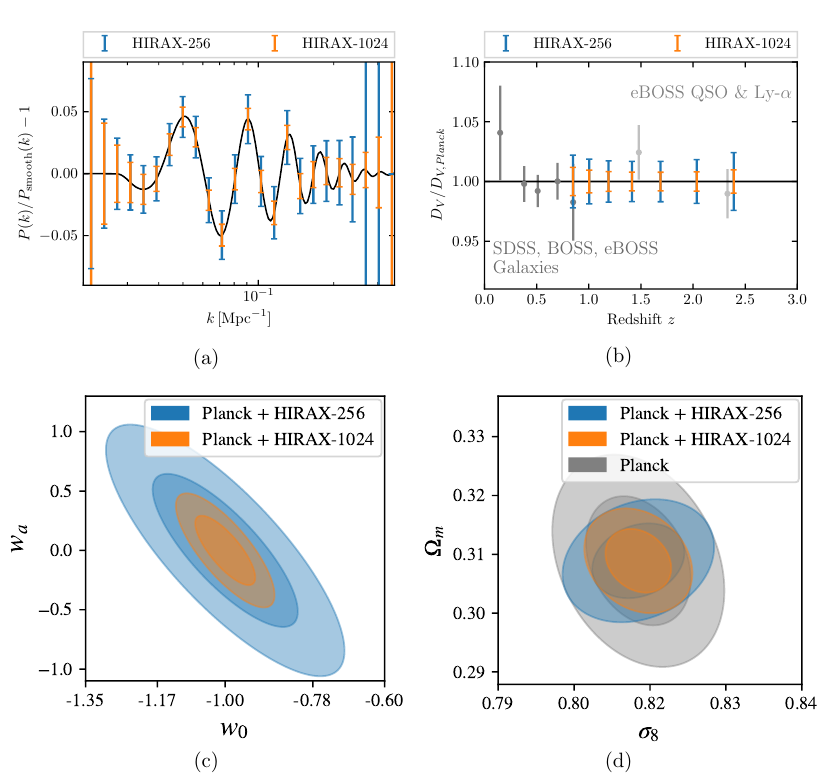}
    \caption{Forecast constraints on
    (a) the BAO feature of the matter power spectrum, 
    (b) distance scale, $D_V$ evolution,
    (c) dark energy equation of state parameters $w_0$ and $w_a$,
    (d) large-scale structure parameters $\sigma_8$ and $\Omega_{M}$,
    for \hirax-256 and \hirax-1024.
    The parameter contours shown represent 68\% and 95\% confidence intervals (shaded and lightly shaded regions respectively). A prior based on \textit{Planck}\cite{2020A&A...641A...6P} results has been added.
    We note that the change in degeneracy direction for the $\Omega_{M}$ and $\sigma_8$ contours between \hirax-256 and \hirax-1024 is due to the different relative contributions of the \textit{Planck} prior to the combined constraints.
    }
    \label{fig:BAO_dv_errors}
\end{figure}

We evaluate the cosmological constraints that can be achieved with \hirax\ using the Fisher forecast method following the approach of Ref.~\citenum{2015ApJ...803...21B}. For reviews on observational approaches to constraining cosmological parameters with BAO measurements, see Refs.~\citenum{2006astro.ph..9591A,2013PhR...530...87W,2015PhRvD..92l3516A}. 
Our analysis is restricted on small scales to linear modes, on large transverse scales by the minimum baseline length, and on large radial scales by filtering of foreground modes below $k_\parallel^{FG} \sim 0.01 \text{ Mpc}^{-1}$. 
In Figure \ref{fig:BAO_dv_errors}(a) we show forecasts for the measurement of the BAO signal in the power spectrum, demonstrating that HIRAX should resolve this feature with high significance. This in turn leads to percent level constraints on the volume-averaged distance measure, $D_{V}(z) = \left[c z D^2_M(z)/H(z)\right]^{1/3}$, 
in several redshift bins as shown in Figure \ref{fig:BAO_dv_errors}(b). Also shown in Figure \ref{fig:BAO_dv_errors}(b) is a compilation of recent BAO constraints from SDSS galaxy, quasar and Lyman-$\alpha$ surveys\cite{2015MNRAS.449..835R,2017MNRAS.470.2617A,2021MNRAS.500..736B,2020MNRAS.498.2492G,2021MNRAS.501.5616D,2021MNRAS.500.3254R,2021MNRAS.500.1201H,2020MNRAS.499..210N,2020ApJ...901..153D}.

We explore cosmological parameter constraints for three different cosmological models, $\Lambda$CDM, $w$CDM and $w_0w_a$CDM. Here $w$CDM allows for variation in a fixed dark energy equation-of-state parameter, $w$, while $w_0w_a$CDM, fits for an evolving equation-of-state $w(a) = w_0 + (1-a)w_a$, as in Ref.~\citenum{2006astro.ph..9591A}.
Marginalized constraints on a subset of the full parameter set are shown in  Table \ref{tab:Cosmo_errors} for the HIRAX 256-element and 1024-element arrays including priors based on constraints from the \textit{Planck} Satellite\cite{2020A&A...641A...6P}. These are compared to current state-of-the-art constraints from the eBOSS cosmology analysis\cite{2021PhRvD.103h3533A}, which are combined with constraints from \textit{Planck}, Pantheon SNe Ia\cite{2018ApJ...859..101S} and DES Y1\cite{2018PhRvD..98d3526A}. 

In Figures \ref{fig:BAO_dv_errors}(c) and \ref{fig:BAO_dv_errors}(d) we show marginalized parameter contours for the dark energy equation-of-state parameters as well as large-scale structure parameters $\Omega_M$ and $\sigma_8,$ respectively, for both the \hirax\ 256-element and 1024-element arrays with \textit{Planck} priors. The computed dark energy figures of merit (FoM\cite{2006astro.ph..9591A}), corresponding to the inverse of the area enclosed by the 68\% confidence contours in the marginalized $w_0-w_a$ constraints, are 60 for \hirax-256 and 360 for \hirax-1024. This is competitive with other planned dark energy experiments\cite{2015ApJ...803...21B}. We note that the \textit{Planck} prior dominates the $\Omega_M$ constraint but that the \hirax\ 1024-element array, in particular, can further constrain $\Omega_M$ and $\sigma_8$.

\begin{table}[ht]
	\centering
	\begin{tabular}{|c|c|c|c|c|}
		\hline
		\text{HIRAX-256} + \textit{Planck}&	$\sigma_8$  & $\Omega_\Lambda$ & $w_0$ & $w_a$ \\
		\hline
		$\Lambda$CDM & 0.0044 & 0.0039 & - &- \\
		\hline
		 $w$CDM & 0.0047 & 0.0042 & 0.0739 & - \\
		 \hline
		 $w_0 w_a$CDM & 0.0053 & 0.0043 & 0.1223& 0.4332 \\
		\hline
		\text{HIRAX-1024} + \textit{Planck}&	 &  & & \\
		\hline
		$\Lambda$CDM & 0.0027 & 0.0034 & - &- \\
		\hline
		 $w$CDM & 0.0028& 0.0036 & 0.0316 & - \\
		 \hline
		 $w_0 w_a$CDM & 0.0038 & 0.0037 & 0.0506& 0.1965 \\
		\hline
		\text{eBOSS} + \textit{Planck} + \text{SNe Ia} + \text{Lens.}& &  &  &  \\
		\hline
		$\Lambda$CDM & 0.0056 & 0.0047 & - &- \\
		\hline
		 $w$CDM & 0.0092& 0.0066 & 0.027 & - \\
		 \hline
		 $w_0 w_a$CDM & 0.0093 & 0.0069 & 0.073& 0.5200 \\
		\hline
	\end{tabular}
	\caption{\label{tab:Cosmo_errors} Marginalized 68\% cosmological parameter forecast constraints for the \hirax\ experiment compared to current eBOSS\cite{2021PhRvD.103h3533A} results for $\Lambda$CDM, $w$CDM and $w_0 w_a$CDM cosmologies.
	}
\end{table}

\subsubsection{Cosmological Analysis Pipeline Status}

We can obtain more realistic forecasts for HIRAX 21~cm power spectrum constraints using detailed telescope simulations. We adopt the $m$-mode approach\cite{2014ApJ...781...57S,Shaw:2014khi,2020PASP..132f2001L}, which is appropriate for a drift-scan instrument such as HIRAX. Decomposing daily scans into $m$-modes along the sidereal time direction allows us to decouple these modes and analyse them independently, which makes the analysis of large arrays computationally tractable. The simulated visibility, ${\bm{v}}={\mathbf{B}}\, \bm{a}+{\bm{n}},$ can be written, for all baselines and frequencies, as a sum of the sky signal, $\bm{a},$ processed by the instrument response or so-called beam transfer matrix, $\mathbf{B},$ and instrument noise, ${\bm{n}}$. The beam transfer matrix encodes information about the telescope beam and pointing, the baseline layout, and the instrument noise. Here, we consider a Gaussian primary beam but directivities estimated from electromagnetic simulations of the HIRAX antenna\cite{2021SPIE11445E..5OS} can also be used in the simulation pipeline.
The input sky model comprises simulated 21cm fluctuations with Gaussian statistics and simulated Galactic and extragalactic foregrounds. Currently, these are generated using Gaussian random field realisations of an input 21~cm power spectrum as well as a Galactic foreground model based on that of Ref.~\citenum{2005ApJ...625..575S} with the large-scale spatial distribution of the signal constrained to be similar to that of the Haslam 408~MHz map\cite{1982A&AS...47....1H}. Future efforts include making use of a more sophisticated signal model based on an empirical \hone\ halo model applied to $n$-body simulations. Given an instrument model and survey specification, the beam transfer matrix can be computed and is used to generate synthetic visibility data. In the results shown here, we use a compact grid of 7$\times$7 dishes (covering the most cosmologically relevant angular scales) and assume a survey conducted over declinations from $-40^\circ$ to $-20^\circ$ through 7 repointings of the array. The noise is scaled such that the redundancy of the baselines of this 7$\times$7 sub-array are equivalent to that of the 1024 element array with a survey length of four months of integration time per simulated fixed-elevation pointing of the array.

The process of estimating a power spectrum from simulated data is
adapted closely from Ref.~\citenum{Shaw:2014khi} but we briefly outline it
here. In the results shown here, the simulated visibilities are
assumed to have negligible calibration residuals but the inclusion of
a realistic calibration pipeline and handling for data excised due to,
e.g., radio-frequency interference, is being developed. The visibility dataset is then compressed by filtering out modes that the HIRAX baselines are insensitive to using singular value decomposition. The next step involves deprojecting galactic and extra-galactic foregrounds, which is the main challenge for a detection of 21cm intensity fluctuations. Foreground removal relies on smooth spectral behaviour of the foregrounds and involves high-pass spectral filtering of the data to retain primarily the high-frequency spectral 21cm modes. Here, we apply a Karhunen-Lo\`eve filter based on an ideal instrument with perfect statistical knowledge of the foregrounds. The power spectrum, binned in $k_\parallel$ and $k_\perp,$ is then estimated from the filtered data.

\begin{figure}[h]
    \centering
    \includegraphics[width=\textwidth]{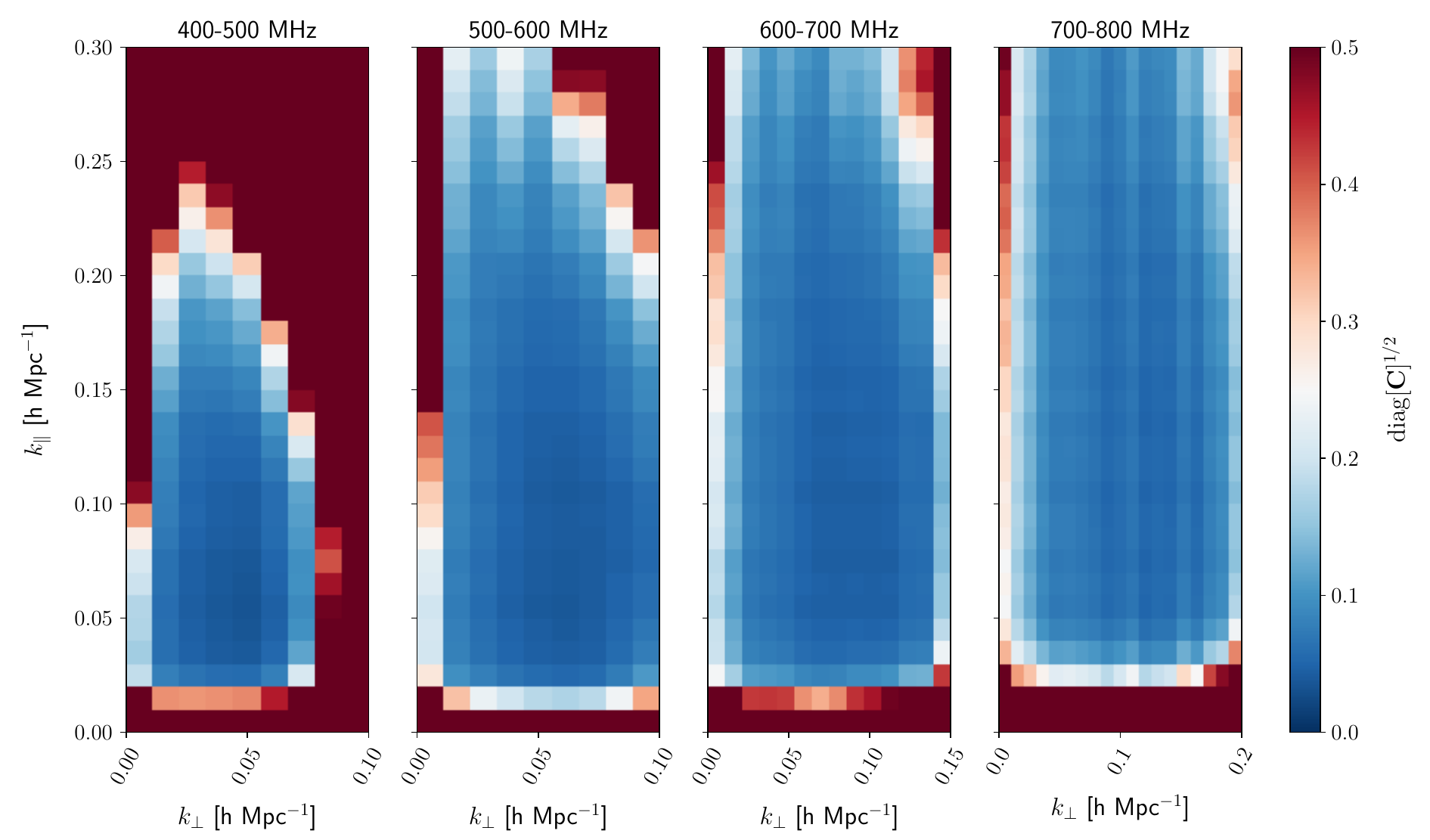}
    \caption{Estimated relative errors in recovered $P(k_{\parallel}, k_{\perp})$ band-powers (from diagonal elements of the derived covariance matrices) over 100~MHz sub-bands spanning the \hirax\ bandwidth using the $m$-mode pipeline. A foreground filter has been applied here assuming ideal knowledge of the instrument.}
    \label{fig:mmode_gaussbeam_unwindowed_allplots}
\end{figure}

The plots in Figure~\ref{fig:mmode_gaussbeam_unwindowed_allplots} show examples of the estimated power spectra constraints using the $m$-mode approach for a Gaussian beam in four different uniform frequency bands. In this case, the mixing matrix for an unwindowed estimator is used. See Ref.~\citenum{2020PASP..132f2001L} for a detailed discussion on different quadratic power spectrum window functions. At the relevant scales of interest as mentioned above and depending on frequency channel considered, we find the binned power spectrum is recovered with relative error from $5\%$ to $8\%$ in the approximate range of  $0.05 < k < 0.10~h~{\rm Mpc}^{-1}$, noting that the set of baselines considered in these simulations also affects these limits. A minimum-variance power spectrum estimator reduces the uncertainties by about a factor of two, at the price of a more complicated correlation structure between the bins. In future work, estimated cosmological constraints will be derived from these power spectrum estimates, and contrasted with the Fisher matrix results in Section~\ref{s:fisher}.

\subsubsection{Systematics Analysis}

We have additionally extended the above analysis to examine the effect of unmodeled systematics on the extraction of accurate power spectra. This involves injecting systematic perturbations to the simulated visibilities by assuming distributions of systematic parameters that affect the primary beam, pointing or array layout of the simulated telescope. For the primary beams, electromagnetic
simulations of the \hirax\ dish-feed systems have been performed\cite{2021SPIE11445E..5OS}, including parameter sweeps of physical tolerances of the system under investigation such as feed positioning relative to the dish. Linear models for the perturbations arising from small changes in these tolerances on the beam directivities are then constructed and propagated to the visibilities which are run through the power spectrum estimation pipeline. For layout-based effects such as errors in dish positioning, a similar approach is used where the linear order perturbation on the visibilities due to these uncertainties are propagated through the pipeline. In both cases care is taken to keep track of systematic correlations between baselines due to sharing systematics from the same dishes. This analysis was used to inform aspects of the system level requirements outlined below, in Section~\ref{s:requirements}.

\subsection{Beamforming and Real-time Analysis}
\label{sec:beamforming}

\hirax\ will sum the channelized baseband data from the array elements
to form beams on the sky for use in the transient/FRB, pulsar, and \hone\ 
absorber searches.  The \hirax\ beamforming will be fully described
in future work, but we provide a brief overview here.

The summing required for the beamforming operation reduces to a 
Fourier transform for a calibrated, planar array. Each location 
of a formed beam on the sky then corresponds to a frequency-dependent $k$
in this Fourier transform. For a close packed array, there are roughly
as many independent beams inside the primary beam as there are
antennas so, for $n$ antennas, and with $n$ beams, the total computational cost becomes
an order $n^2$ matrix multiplication. The visibilities already cost $n^2$
to generate, and so this brute force beamforming at worst increases the
computational burden on the correlator by a constant, order unity
factor. For regularly spaced antennas, Fast Fourier transforms (FFTs) can be
used at single frequencies to potentially significantly speed up this process.
However, FFTs naturally pick out integer values of $k$ and therefore require 
additional steps to align the beams across frequencies, which is important for
detecting dispersed signals. Since modern GPUs are sufficiently performant at the
low bit-depth matrix multiplications required by the brute force beamforming, we find this approach
attractive for a 256-element array. For example, for 4-bit operations, 
the NVIDIA RTX 3090 as well as the NVIDIA A40s in use in the HIRAX-256 X-engine 
are capable of over 500 tera operations per second (TOPS).  If this processing rate 
can be achieved in practice, a single GPU has enough processing power to form 256
beams from 256 HIRAX antennas. Preliminary GPU kernels have been developed
and found to perform within a factor of two of the theoretical limit.  

With the decision to use a brute force beamformer, we can now describe
how the \hirax\ beamforming will work.  The beam-crossing time is
relatively short for \hirax\ ($\sim$1 minute) 
and so we will use tracking beams that follow sources as they transit
the primary beam.  The RMS per-antenna errors introduced by using
4-bit arithmetic are $\sim$3$^\circ$ with an RMS amplitude fractional error
of $0.035$.  Since the data from the F-engine arrive as 4-bit
integers, and the errors introduced by 4-bit beamforming are small, we
plan to use 4-bit arithmetic to do the beamforming.  The beamforming phases
will be updated approximately every second.  We plan to beamform the intensity
data only, using both $XX$ and $YY$ polarization, and do not plan to
form real-time cross-polar beams.  Since many scientific uses of the
formed beams require higher frequency resolution than that provided by
the F-engine, the formed beams will be upchannelized\footnote{Resampling at higher frequency resolution and hence number of channels.} (see, e.g., Refs.~\citenum{2018ApJ...863...48C,2021ApJS..255....5A})
before being squared and summed to produce high frequency-resolution
power beams.  These upchannelized beams can then either be integrated
in the X-engine for the 21cm absorber search, or passed to the FRB
backend for the FRB search.  We additionally will send a small number
($\sim$10) of non-squared beams to the pulsar backend for the pulsar
search.  This gives the pulsar backend the option of carrying out
coherent de-dispersion for millisecond pulsar searches. The utility of
\hirax-256 for pulsar studies is still under study but we expand upon
the planned FRB survey below.

\subsubsection{Fast Radio Burst Survey}
The \hirax\ fast radio burst (FRB) survey will use the search pipeline
developed for \chime\cite{2018ApJ...863...48C}.  Briefly, the formed,
upchannelized beams will be sent to a separate search engine, where
they will be searched for transient events incoherently (for efficiency).  The search
will be carried out by a computationally efficient tree-dedispersion
algorithm.  Detected FRBs will be flagged, and baseband data will be
written to disk for off-line analysis.  The instantaneous detection
threshold will be around 10\,$\sigma$, though the exact threshold will
need to be tuned on-site and will depend on the false positive rate
from local RFI. At 256 dishes, \hirax\ has
similar collecting area and sensitivity to \chime, and so we expect the
redshift and DM distributions will be similar to the \chime\ FRB
catalog\cite{2021arXiv210604352T}.  With a smaller field of view, we
expect an event rate roughly half of \chime's, or about 1 FRB per
day.  

We will build a minimum of two outrigger stations of 8--16 dishes each in order to localize
FRBs detected by \hirax\ to sub-arcsecond accuracy.  The central goal is
to localize the FRBs to within a host galaxy for the large majority of
\hirax\ FRBs.  Additionally, with $\sim$1000 km baselines, we hope to
localize FRBs to tens of milliarcseconds (mas) for studies of FRB environments within
their host galaxies.  Bandwidth
constraints prevent us from continuously streaming baseband data, so
the outriggers will store 1-2 minutes of baseband data in memory (45
GB/s/dish required).  When the core detects an FRB, it will send a
trigger to the outriggers that will then write their baseband data to
disk, for off-line correlation.  Post-processing
of the core data will typically result in a higher SNR than the
initial detections.  Reasons include the ability to reform a beam directly on the
FRB, coherently de-disperse the data, and expend more effort on
RFI mitigation.  We expect 15\,$\sigma$ to be typical for the
post-processed SNR.  We target 5\,$\sigma$ detections between the core
and each outrigger station, which results in outrigger station sizes
of 16-24 dishes.  At a fixed false-positive rate, there is
an increase required in the SNR due to the possibly large phase space
over which one is searching for possible FRBs.  For 2-dimensional
searches using narrow-band, very long baseline interferometry data, this factor can be large.
However, since we are targeting a detection by each station when
correlated against the core, the result is a single timing offset from
broad-band data, so calculations show the phase-space effect is small
for \hirax.  Further, the bulk of the expected false positives arise
from small enough timing errors that we would still localize the FRB
to the correct galaxy.  

Localizations will be fundamentally tied to astronomical calibrators,
and we plan to be able to calibrate using observations up to $\sim$ 1
hour after the FRBs occur.  We have demonstrated moderately-priced
off-the-shelf clock systems 
that are capable of holding the sub-ns timing accuracy we need on multi-hour
timescales\cite{mena_parra_clocks}.  At \hirax\ frequencies,
ionospheric corrections are important, and we plan to use a
combination of astrometric calibrators, GNSS satellites, and real-time
publicly available ionospheric maps.  The worst-case scenario of using
publicly available ionosphere maps/models corresponds to a typical
uncertainty of a few tenths of a TEC unit,\footnote{The total electron content or TEC of the ionosphere is reported in units of $10^{16}
  e^{-}/m^2$.  1 TEC unit corresponds to a DM of $3.3 \times
  10^{-7}$.} which translates to roughly 1 nanosecond.  For outrigger
separations of $\sim 1000$ km, this corresponds to a 60 mas error, and
so we do not expect the ionosphere to limit our ability to assign FRBs
to host galaxies.

\section{Instrument Requirements}\label{s:requirements}

With the science requirements and analysis methods described in
\S\ref{s:analysis}, we can now define the following high-level system requirements for the \hirax\ array.  The system requirements are derived by propagating the higher level science requirements, starting from the top-level requirement on the dark energy FoM, to lower level system requirements based on the simulations methodology described in \S\ref{s:analysis}.
The survey and array layout system requirements were described in \S\ref{s:survey}, and here we focus on the requirements for the ensemble of telescopes comprising the array.  Since there are typically multiple system requirements that are derived from any given individual science requirement, we consider tradeoffs in tolerances for the system requirements to allocate the error budget.
Across the \hirax\ array, we find that the gains must be stable to 1\% over 1-minute integrations across the full bandwidth, relative bandwidths must be stable to 0.75\% over the beam crossing time, and the tolerance on the primary beam FWHM must be 0.05\% across the full bandwidth. In addition, the dishes should have an aperture efficiency of 0.7, the antenna cross-talk should contribute an excess noise level of no more than 10~K on the two shortest dish separations, and the antenna spill should have a maximum noise level of 5~K at 400~MHz, increasing to 10~K at 800~MHz. Finally, telescope pointing and positioning must be repeatable across the array: we require that the dish boresight vectors must be parallel within $5'$~RMS,
the foci of all dishes must lie in a plane with $<5$~mm RMS deviation orthogonal to the plane, and the foci must form a regularly spaced grid with separation distances precise to $<2.5$~mm RMS. We note that the analyses used to derive these constraints were based on science targets for the 1024-element array as the requirements will be fixed across the expected array deployments.

The system requirements for the array as a whole form the foundation
for defining more specific requirements for the \hirax\ telescope
mechanical structure, comprising the dish, mount, and receiver support
structure.  In total, we have defined over 60 specifications that are
discussed in detail in the \hirax\ telescope mechanical assembly
requirements document, which is available
online.\footnote{\url{https://hirax.ukzn.ac.za/wp-content/uploads/2021/12/HIRAX_REQ001_002_V1_Baselined-signed.pdf}}
Broadly speaking, tolerances related to redundancy fall into two
general categories: pointing or positioning error, and beam error.
Pointing and positioning errors are analyzed with a simple geometric
model of the telescope to compute the error stackup.  Starting with
the requirements for boresight vector parallelism and foci positioning
within a plane, the error budget is allocated to various telescope
sub-components in accordance with the mechanical difficulty of meeting
each target.  The telescope parameters that are governed by this
geometric error stackup include the dish vertex position relative to
the elevation axis, orthogonality of the dish boresight vector and
elevation axis, position and alignment angle of the elevation axis
within the array, and the elevation pointing angle.  Errors in the
beam shape are assessed with electromagnetic simulations of the dish,
receiver, and receiver support structure.  We use CST Studio
Suite\footnote{\url{https://www.3ds.com/products-services/simulia/products/cst-studio-suite/}}
to simulate the impact of the receiver position with respect to the
dish focus, receiver orientation angle relative to the boresight
vector, and deviations in the dish surface with respect to the average
best-fit paraboloid.  Table~\ref{t:mech_precision} summarizes the
target precision values for the most important telescope mechanical
parameters.

\begin{table}
\centering
\begin{tabular}[pos]{ll}
\hline
{\bf Telescope mechanical parameter} & {\bf Target precision (RMS)} \\
\hline
Receiver position relative to focus & 0.5~mm \\
Receiver orientation relative to boresight vector & $2.5'$ polar and azimuthal \\
Dish surface deviations & 1~mm \\
Dish vertex position relative to elevation axis & 1~mm \\
Orthogonality of boresight vector and elevation axis & $1'$ \\
Elevation axis position within the array & 0.5~mm in array plane \\
                                         & 1~mm out of array plane \\
Elevation axis alignment within the array & $1'$ \\
Elevation pointing angle & $1'$ \\
\hline
\end{tabular}
\vskip 5pt
\caption{Target precision values for \hirax\ telescope mechanical
  structure}\label{t:mech_precision}
\end{table}

\section{Current Status and Conclusions}\label{s:conclusions}

The design of most aspects of the 256-element \hirax\ deployment have been finalised and site development is expected to commence in 2022.
Currently, multiple hardware testing activities are underway, with instrumented prototype $f/0.38$ and $f/0.25$ dishes deployed at \hartrao\ for RF front-end characterisation, and parallel efforts in progress at
Dominion Radio Astrophysical Observatory and the Green Bank Observatory. Beam characterisation analyses using drone-based and holographic measurements are being carried out.
The design of the F- and X-engine systems are complete and the 256-element X-engine is assembled, with one node deployed at the Bleien Observatory in Aargau, Switzerland, for trial operation using on-sky data. A significant milestone has been reached in the finalisation of the specification for the telescope mechanical system, with the expectation that initial deployed dishes making use of the final design should be assembled for testing in 2022.

HIRAX-256 will be a powerful instrument for 21~cm cosmology, radio transient, and \hone\ absorber science. While much emphasis has been placed on the control of systematics from \textit{a priori} studies, we expect the analysis of the initial on-sky data to be extremely informative for progressing the analysis techniques in use as well as the design of future 21~cm experiments.

% \subsection*{Disclosures}
% Conflicts of interest should be declared under a "Disclosures" header. If the authors have no relevant financial interests in the manuscript and no other potential conflicts of interest to disclose, a statement to this effect should also be included in the manuscript.

\subsection*{Acknowledgments}

The authors would like to thank the anonymous reviewers for their helpful comments on the draft. This work is based on the research supported in part by the National Research Foundation of South Africa (Grant Numbers 98772, 128919, 107797, 120809, 120700) and in part by grant numbers 200021\_192243, 200020\_182231, IZLSZ2\_170907, 20FL21\_186180 and 20FL20\_201479 from the Swiss National Science Foundation.
AR, DC, TV, CB, CF, MK, VN, AVS, ET, and JPK  acknowledge funding by the Swiss National Science Foundation. 
KM acknowledges support from the National Research Foundation of South Africa.
AW and SB acknowledge support from the South African Research Chairs Initiative of the Department of Science and Technology and National Research Foundation and from a VC 2020 Future Leaders award. 
RM, DK, LR, DdV and CP are supported by the South African Radio Astronomy Observatory (SARAO) and the National Research Foundation (Grant No. 75415 and 75322) and RM is also supported by the UK STFC Consolidated Grant ST/S000550/1.
WN and DC acknowledge the financial assistance of the South African Radio Astronomy Observatory (SARAO) towards this research (www.sarao.ac.za). LN, MH, AP, WT acknowledge support from the National Science Foundation under Grant No. 1751763. EK acknowledges support by a NASA Space Technology Research Fellowship. TCC and PB  acknowledge support from the NASA Jet Propulsion Laboratory Strategic  R\&TD awards. Part  of  this work  was  done  at  Jet  Propulsion  Laboratory,  California Institute of Technology,  under a contract with the National Aeronautics and Space Administration. The research of OS is supported by the South African Research Chairs Initiative of the Department of Science and Technology and National Research Foundation.
HCC acknowledges the support of the Natural Sciences and Engineering Research Council of Canada (NSERC;  RGPIN-2019-04506).
ADH acknowledges support from the Sutton Family Chair in Science, Christianity and Cultures and from the Faculty of Arts and Science, University of Toronto.
AZ was supported by a University of Toronto Excellence Award.
MGS, SP and JT acknowledge support from the South African Radio Astronomy Observatory (SARAO)  
and National Research Foundation (Grant No. 84156). 
Computations were performed on Hippo at the University of KwaZulu-Natal, on the Baobab cluster at the University of Geneva and on the Niagara supercomputer at the SciNet HPC Consortium. SciNet is funded by: the Canada Foundation for Innovation; the Government of Ontario; Ontario Research Fund - Research Excellence; and the University of Toronto\cite{Loken_2010,10.1145/3332186.3332195}. This work made use of the \texttt{numpy}\cite{2020NumPy-Array}, \texttt{scipy}\cite{2020SciPy-NMeth}, \texttt{matplotlib}\cite{mpl}, and \texttt{astropy}\cite{astropy:2013, astropy:2018} software packages.

% \subsection* {Code, Data, and Materials Availability} 
% As relevant, declare the availability of computer software code, data, and/or materials used in the research results reported in the manuscript. Provide specific access information or restrictions for code, data, and materials (i.e., links to repository access addresses, and/or guidance on commercial or public access). Note: reporting in this section is required for the \textit{Journal of Biomedical Optics} and \textit{Neurophotonics}. 

%%%%% References %%%%%

\bibliography{hirax_proc}   % bibliography data in report.bib
\bibliographystyle{spiejour}  % makes bibtex use spiejour.bst

\listoffigures
\listoftables

\end{spacing}
\end{document}